\documentclass[12pt,preprint]{aastex}

\usepackage{CJK}
\begin{document}
\begin{CJK}{UTF8}{gbsn}

\title {LAMBDA BOO ABUNDANCE PATTERNS: ACCRETION 
FROM ORBITING SOURCES}

\author{M. Jura\altaffilmark{a}}
\altaffiltext{a}{Department of Physics and Astronomy, University of California, Los Angeles CA 90095-1562; jura@astro.ucla.edu}

\begin{abstract}
The abundance anomalies in ${\lambda}$ Boo stars are popularly explained by element-specific mass inflows at rates that  are much greater than empirically-inferred bounds for interstellar accretion.  Therefore,  a ${\lambda}$ Boo star's thin outer envelope must derive from  a companion star, planet, analogs to Kuiper Belt Objects or a circumstellar disk.  Because radiation pressure on gas-phase ions might selectively allow the accretion of carbon, nitrogen, and oxygen and inhibit the inflow of elements such as iron, the source of the acquired matter need not contain dust.  We propose that 
at least some  ${\lambda}$ Boo stars   accrete 
from the winds of  hot Jupiters.  
  \end{abstract}

\section{INTRODUCTION}
Approximately 2\% of main sequence A-type stars are similar to ${\lambda}$ Boo in having essentially solar carbon, nitrogen
and oxygen abundances but subsolar abundances of heavier elements such as iron \citep{Paunzen1991,Gray1992}.  A plausible
 model to explain these distinctive abundances is that these stars have accreted gas but not dust
\citep{Venn1990}.  In this scenario, radiation pressure from the star prevents grains with
their high opacity from accreting while gas is not so inhibited \citep{Kamp2002a,Martinez-Galarza2009}.  

Although the general picture that ${\lambda}$ Boo stars acquired their distinctive abundances by selective accretion is promising, there are objections which must be resolved \citep{Cowley2014}.
 For example, most ${\lambda}$ Boo stars do not exhibit an infrared excess \citep{King1994, Paunzen2003}.  Consequently, there is no evidence for either circumstellar dust or nearby interstellar dust.  Also, main-sequence A-type stars with dust do not always 
 exhibit ${\lambda}$ Boo photospheric abundances  \citep{Kamp2002b}.
 Seemingly, unless there is undetected dust, some  additional mechanism to explain selective elemental accretion should be considered.
 
  The popular model for ${\lambda}$ Boo stars requires  a substantial accretion rate.   In particular,
 depending upon the amount of deep mixing and meridional circulation, required accretion rates range between 6 ${\times}$ 10$^{11}$ g s$^{-1}$ and 6 ${\times}$ 10$^{14}$ g s$^{-1}$ \citep{Turcotte1993,Turcotte2002}, comparable to the solar wind outflow of ${\sim}$10$^{12}$ g s$^{-1}$.
 As described in detail below,  these values are orders of magnitude  greater  than empirically-determined bounds on accretion rates by stars  
from the  interstellar medium.  Therefore, the thin outer envelope must be acquired  from an orbiting source close to the star.

 A process that has been invoked to explain the anomalous circumstellar abundances of  ${\beta}$ Pic  \citep{Fernandez2006, Xie2013} but not yet considered for ${\lambda}$ Boo stars is radiation pressure
on individual ions.  This effect is highly element specific  and therefore may be important in explaining the abundance pattern in ${\lambda}$ Boo stars.
If so, then dust need not be a constituent in the reservoir of the accreted material,  allowing us to consider  sources for accretion onto ${\lambda}$ Boo stars that were
previously neglected.

 In Section 2, we examine and discard the possibility of interstellar accretion.     In Section 3, we assess the possibility that radiation pressure on atoms and ions can act to allow for selective
accretion with the  characteristic pattern seen in ${\lambda}$ Boo stars.  In section 4, we  assess various possible sources of the matter accreted onto ${\lambda}$ Boo stars and suggest
that one unappreciated possibility is the
 wind from an irradiated hot Jupiter.   We discuss our results in Section 5 and our conclusions in Section 6.
 
 \section{INTERSTELLAR ACCRETION?}
 
 \citet{Martinez-Galarza2009} used Bondi-Hoyle theory to compute interstellar accretion onto ${\lambda}$ Boo stars.
   Because the applicability of this approach is uncertain,  \citep{Koester1976,Farihi2010},
empirical measures of accretion rates   should be considered.  
A sensitive upper bound on the time-averaged rates of interstellar accretion onto stars can be derived from the amount of
hydrogen found in the outer mixing zones of DB white dwarfs, stars where the dominant element in the atmosphere is helium.  Because much or all of this hydrogen either had a circumstellar origin or was primordial \citep{Bergeron2011},  only upper bounds can be derived for the rate of interstellar accretion,  $dM/dt$.  In a catalog
of 57 DBs within 80 pc of the Sun, the time-averaged value of $dM/dt$ typically is less
than or equal to 10$^{6}$ g s$^{-1}$ \citep{Jura2012}.     

For stars of mass, $M_{*}$, the Bondi-Hoyle accretion rate varies as $M_{*}^{+2}$ .  White
dwarf stars typically have 1/3 the mass of a ${\lambda}$ Boo star \citep{Kleinman2013,Paunzen2002}, and therefore their accretion rates from the interstellar
medium might be a factor of 10 lower.  However, even making this adjustment, the time-averaged
accretion rate inferred for DB white dwarfs is nearly a factor of 10$^{5}$ lower than the minimum required rate.  Because ${\sim}$50\% of the 34 ${\lambda}$ Boo stars discussed by \citet{Heiter2002b} lie within 80 pc of the Sun and therefore within the local interstellar bubble \citep{Lallement2003}, this comparison with nearby white
dwarfs is meaningful. 

 We also consider empirical measures
of ``instantaneous" accretion.    We therefore use measures of accretion rates onto DA white dwarfs (stars where the atmosphere is dominated by hydrogen) that are warmer than 13,000 K  where the settling time of heavy elements below the photosphere is only days \citep{Koester2009}.  In a sample of 87 such stars,
the maximum carbon accretion rate is 2 ${\times}$ 10$^{6}$ g s$^{-1}$ \citep{Koester2009}.  Because solar abundances \citep{Lodders2003} are assumed for the source material, ${\lambda}$ Boo
stars are required to accrete carbon at a rate of 2 ${\times}$ 10$^{9}$ g $^{-1}$, a factor of 1000 greater than the maximum value measured for white dwarfs.  The true discrepancy is probably substantially  larger because the most likely source of the white dwarf's photospheric carbon is circumstellar rather  than interstellar \citep{Jura2014}.    We conclude that  empirical bounds on interstellar accretion rates are much
lower than what is required to explain ${\lambda}$ Boo stars.

\section{ACCRETION GOVERNED BY RADIATION PRESSURE}

Before exploring the different possible sources of accretion onto ${\lambda}$ Boo stars, we  consider
an additional element-sorting mechanism  for sources that are dust-free.
Specifically, we extend models invoking radiation pressure on individual ions developed to explain the distinctive composition of  ${\beta}$ Pic's circumstellar gas to ${\lambda}$ Boo stars.
 
 Following
notational convention, we define ${\beta}$ as the ratio of the outward force of radiation pressure compared to the inner
force of gravity.   If the star has radius, $R_{*}$, effective temperature, $T_{*}$, mass, $M_{*}$ and emergent flux at the stellar surface, $F_{\nu}$, then:
\begin{equation}
{\beta}\;=\;\left(\frac{{\pi}\,e^{2}}{m_{e}\,c^{2}}{\sum}_{j}\,f_{j}\,F_{\nu}\right)\left(g\,m_{i}\right)^{-1}.
\end{equation}
Here,  $e$ denotes the charge of an electron of mass, $m_{e}$, $c$ is the speed of light, $g$ is the gravitational acceleration at the photosphere and $m_{i}$ is the mass of the $i$'th ion with resonance transitions of oscillator strength, $f_{j}$.  We consider environments sufficiently far from the host star that ions mostly lie in their ground states.

Qualitatively, using the Wien approximation to the Planck curve and assuming that in the ultraviolet the opacity is so high that we observe only the upper layers of the atmosphere  and that the photospheric lines are weak, we  write that:
 \begin{equation}
F_{\nu}\;{\sim}\;\frac{2\,{\pi}\,h{\nu}^{3}}{c^{2}}\,e^{-h{\nu}/kT_{min}}
\end{equation}
where $T_{min}$ is the temperature minimum in the photosphere or ${\sim}$ 0.75 $T_{*}$ \citep{Kurucz1979}.
A low value of ${\beta}$ requires a low value of $F_{\nu}$ which requires that the resonance lines lie at relatively short wavelengths.  
Important lines for C II , N I and O I
lie at 1335 {\AA},  1200 {\AA} and  1302 {\AA}, respectively, while, in contrast,  important lines for Mg, Ca and Fe lie at 2802 {\AA}, 3933 {\AA} and 2599 {\AA}, respectively \citep{Morton1991}.  Therefore, C, N and O can be accreted while Mg, Ca and Fe cannot.  This
qualitative expectation is  borne out by 
detailed calculations \citep{Fernandez2006} for a model of ${\beta}$ Pic  with assumed values of $T_{*}$ = 8000 K and $\log$ $g$ = 4.2,
stellar parameters  appropriate for many ${\lambda}$ Boo stars \citep{Heiter2002b}.   We show in  Figure 1 a plot for ${\lambda}$ Boo of an element's abundance, compared to its solar abundance, $X$, vs.
the relative importance of outward radiation pressure, ${\beta}$, where:
\begin{equation}
X\;=\; \log \left(\frac{n(X)}{n(H)}\right)_{*}-\;\log \left(\frac{n(X)}{n(H)}\right)_{\odot}
\end{equation}
There are two classes of elements: those with high values of $X$ and low values of ${\beta}$ in their likely dominant state of ionization and those with low values of $X$ and
high values of ${\beta}$.

This analysis  extend to elements beyond those considered in Figure 1.  Both H and S also are computed to
have low values of ${\beta}$ and can be accreted while Si and Mn have high values of ${\beta}$  \citep{Fernandez2006} and would be inhibited from accreting, consistent
with the general pattern of abundances among ${\lambda}$ Boo stars \citep{Heiter2002}.   

One complication to this picture is that there are some
elements, such as Na, where ${\beta}$ can be high in one state of ionization and low in another.  For such an element, the net balance among all the potential
states must be determined in order to assess whether it is accreted.  Although there are uncertainties,
we argue that  outward radiation pressure on individual elements might play a key role in explaining the abundance pattern of ${\lambda}$ Boo stars.  
\begin{figure}
 \plotone{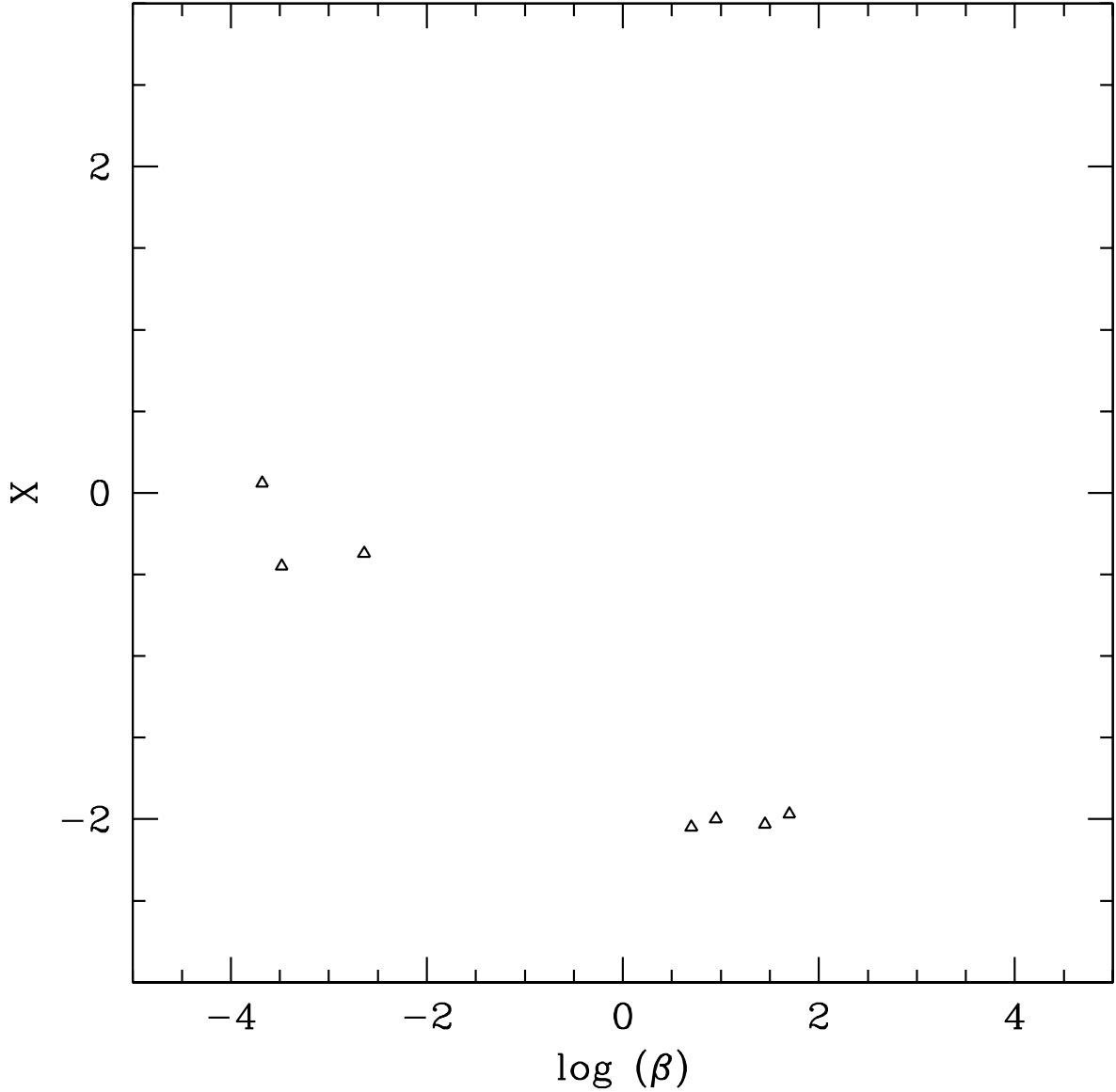}
\caption{Plot of abundance relative to solar \citep{Venn1990}  defined by $X$ in Equation (3) vs. ${\beta}$ defined in Equation (1) for different elements in
their likely dominant state of ionization \citep{Fernandez2006}: C II, N I, O I, Fe II, Ca II, Ti II, Mg II.  The three elements with approximately solar abundances are C, N and O.  A clear distinction  in values of $X$ between those with high ${\beta}$ and those with low ${\beta}$ 
is evident.} 
\end{figure}

\section{ORBITING SOURCES}

The alternative to interstellar accretion is accretion from some material gravitationally bound to the host star.  
We now assess different possibilities.

\subsection{Companion Stars?}

A possible scenario to explain ${\lambda}$ Boo abundances is the accretion of the outflow from a  stellar companion.
In fact, there are rare instances of spectroscopic binary systems where two stars orbit close to each other and one star exhibits a ${\lambda}$ Boo abundance pattern \citep{Narusawa2006, Waters1992}. However, such situations probably do not
explain most ${\lambda}$ Boo stars \citep{Stutz2006,Griffin2012}.  A binary configuration of two more distant stars which
might account for some ${\lambda}$ Boo objects would be an A-type star accreting from the wind of a late type companion. The typical
mass loss rate from a late-type main-sequence star is ${\sim}$ 10$^{12}$ g s$^{-1}$ \citep{Wood2002}, comparable to the mass accretion rate required to explain ${\lambda}$ Boo stars.  Because the characteristic wind speeds from late type stars likely are near their escape speed of 600 km s$^{-1}$, only a small fraction of the matter lost from the companion
can be accreted by the ${\lambda}$ Boo star \citep{Debes2006}.  Therefore, accretion of  winds from late-type stellar companions seems unlikely to explain  most ${\lambda}$ Boo stars.  

\subsection{Rocky Planets or Asteroids?}
The carbon to oxygen ratio measured in the atmospheres of  ${\lambda}$ Boo stars 
typically is within a factor of three of the solar ratio \citep{Heiter2002b}.  In contrast,  most matter from rocky minor planets accreted onto white dwarfs has a carbon to oxygen
ratio that is typically  a factor of  100 lower than solar \citep{Jura2014}.    This marked deficiency of carbon within extrasolar  asteroids -- the building blocks of rocky planets -- argues against their being a  major reservoir of matter accreted onto
${\lambda}$ Boo stars.   

\subsection{Kuiper Belt Analogs or Comets?}

Another source of accreted material might be analogs to Kuiper Belt Objects or comets
 with carbon to oxygen abundance ratios that might be nearly solar  \citep{Jura2015} although perhaps not \citep{Wilson2015}.  \citet{Zuckerman2012} have proposed that colliding comets explain the observed gas
 around 49 Cet, a young main sequence A-type star.  The suggested
 CO production rate is greater than ${\times}$ 10$^{13}$ g s$^{-1}$, large enough to induce
${\lambda}$ Boo abundances.  However, there are difficulties with this model.  It is not clear how matter initially at ${\geq}$100 AU
ultimately accretes onto the host star.  Also, comets  by consisting largely of water, may have very subsolar hydrogen to oxygen ratios.   Accretion of such material
might produce an oxygen overabundance which is not observed.  Finally, 49 Ceti has an infrared excess while most ${\lambda}$ Boo stars do not.  Given
these objections, alternative reservoirs for the accretion onto ${\lambda}$ Boo stars should be considered.

\subsection{Pre-Main-Sequence Disks?}

In the absence of accretion, mixing erases the abundance anomalies in ${\sim}$10$^{6}$ yr \citep{Turcotte1993}.  While accretion from a primordial
disk may lead to the ${\lambda}$ Boo abundance pattern measured  in numerous pre-main-sequence stars \citep{Folsom2014,Cowley2014b}, there are many
${\lambda}$ Boo stars that are sufficiently old that their photospheric abundances are not a relic from the era of  their formation  \citep{Paunzen2002}.   A recent source
of accretion should be considered.

\subsection{Hot Jupiters?}

Because the source must be substantial and likely located relatively near the star to guarantee ultimate accretion, 
we propose that the source of matter accreted onto many ${\lambda}$ Boo stars is the wind from a close-in planet.
As an example, consider WASP 33 (or HD 15082), an A5 main-sequence star orbited by a hot Jupiter with a period
of 1.22 d \citep{Collier-Cameron2010}.  The mass loss rate from this planet may be 2 ${\times}$ 10$^{13}$ g s$^{-1}$ \citep{Bourrier2015},
 sufficiently high to explain  ${\lambda}$ Boo stars.  

Here, we assume that all the mass lost from the planet ultimately is acquired by the star.  Furthermore, we also assume  that individual ions can separate from the general flow as supported by a toy model not presented here.  Detailed calculations are required to assess these requirements.  For example, in the case of the outer circumstellar
gas orbiting ${\beta}$ Pic, the effectiveness of the separation between different elements depends upon the viscosity in the gas \citep{Xie2013}.

In models where the upper atmosphere is heated by ultraviolet and X-ray photons, the wind rate, ${\dot M}$, from a planet at distance $D$ from the can star can be estimated as:
\begin{equation}
{\dot M}\;=\;\frac{3\,{\eta}}{16\,{\pi}\,G{\rho}_{p}}\,\frac{L_{EUV}}{D^{2}},
\end{equation}
where $G$ is the gravitational constant, $L_{EUV}$  is the total stellar EUV luminosity  in the spectral range  ${\lambda}$ $<$ 912 {\AA},  ${\rho}_{p}$ is the planet's
 mean density, and
 ${\eta}$ is an efficiency  factor to describe the conversion of incident radiative energy to kinetic energy
in the outflowing wind \citep{Murray-Clay2009}.  This formula may overestimate the planetary mass loss
rate if the planet's magnetic  field is sufficiently large \citep{Owen2014}.  

Now consider  ``optimistic" values of the different parameters in Equation (4).  We take $R_{p}$ of and ${\rho}_{p}$ for WASP 17b of 1.3 ${\times}$ 10$^{11}$ cm and 0.062 g cm$^{-3}$, respectively \citep{Anderson2011}, the most inflated hot Jupiter currently known \citep{Bento2014}.  We assume $D$ = 6.0 ${\times}$ 10$^{11}$ cm in order for the planet to be stable against tidal disruption \citep{Paczynski1971}.
We also take 
${\eta}$ = 1.  Because of attenuation in the interstellar medium,
the EUV luminosity of most stars is unobservable  and  therefore highly uncertain.  Indirect estimates of $L_{UV}$ can be derived from observed subcoronal emission lines characteristic of gas with temperatures between 50,000 K and 300,000 K.  Such lines 
have been detected in stars with $T$ $<$ 8200 K although  stars as hot as  8800 K may possess coronae \citep{Simon2002}.  Further, approximately 10-15\% of main sequence A-type stars are detected X-ray sources; the underlying physical explanation is not known \citep{Schroder2007}.   Here, to be specific, we consider for comparison the well studied star, ${\beta}$ Pic, with $T$ = 8200 K ${\pm}$ 150 K \citep {Lanz1995}, a representative temperature for many ${\lambda}$ Boo stars.  For ${\beta}$ Pic,
the measured X-ray luminosity between 0.2 and 20 keV is 3 ${\times}$ 10$^{26}$ erg s$^{-1}$, and the extrapolated X-ray luminosity between
0.06 and 5 keV is 3 ${\times}$ 10$^{27}$ erg s$^{-1}$ \citep{Gunther2012}.  There must also be emitted photons with energies between 0.0136 and 0.06 keV.  According to \citet{Bouret2002}, the total chromospheric emission of ${\beta}$ Pic is at least 9.9 ${\times}$ 10$^{28}$
erg s$^{-1}$. While much of this energy is emitted   with ${\lambda}$ $>$ 912 {\AA}, an appreciable fraction is thought to be emitted at ${\lambda}$ $<$ 912 {\AA}.  Here, recognizing that there is great uncertainty, we adopt $L_{EUV}$ = 2 ${\times}$ 10$^{28}$ erg s$^{-1}$ or $L_{X}$/L$_{bol}$ ${\sim}$ 5 ${\times}$ 10$^{-7}$.   If so, then
${\dot M}$ ${\approx}$ 10$^{12}$ g s$^{-1}$, sufficiently large to explain ${\lambda}$ Boo  abundances.

One complication  is that a hot Jupiter may have a sufficiently short period that it can induce activity on its host star.  \citet{Shkolnik2005} have
reported evidence to support ${\sim}$10$^{27}$ erg s$^{-1}$ of such dissipation.  Accordingly, through magnetically controlled interactions,
the host planet might lose 10$^{12}$ g s$^{-1}$ \citep{Lanza2013}.  However, it is uncertain whether models developed for later type stars apply to A-type stars, but
if they do, then the planetary outflow would be sufficiently great to explain many ${\lambda}$ Boo abundances.

This model might apply to to HR 8799, a main-sequence A-type star  orbited by planets in wide orbits \citep{Marois2010} which is also is a ``mild" ${\lambda}$ Boo star \citep{Gray1999}.
Because HR 8799 emits 1.3 ${\times}$ 10$^{28}$ erg s$^{-1}$ in the 0.2 - 2 keV band \citep{Robrade2010}, a wind from a hot Jupiter -- an additional planet interior to  those already known -
might account for the photospheric abundances.

In this analysis, we assume that hot Jupiters have an atmosphere with nearly a solar composition.  Although this result is  unestablished \citep{Burrows2014}, at least Na and K
have been detected in the outer atmospheres of hot Jupiters \citep{Charbonneau2002, Sing2015}.  Further,  winds from hot Jupiters display  evidence for  Mg \citep{Fossati2010, Haswell2012, Vidal-Madjar2013} and possibly Si \citep{Linsky2010,  Ballester2015}.  Current data are consistent with the hypothesis that
the abundances in the atmospheres of hot Jupiters are similar to the material accreted on  ${\lambda}$ Boo stars.

\section{DISCUSSION}

Winds from hot Jupiters may be the source of accretion for many  ${\lambda}$ Boo stars.  For F,G and K stars, the frequency of hot Jupiters is 1.2\% ${\pm}$ 0.38\% \citep{Wright2012}.  Although
the frequency of hot Jupiters orbiting main-sequence A-type stars is unknown, the fraction of stars with giant planets appears to  peak for stellar masses near 1.9 M$_{\odot}$ \citep{Reffert2015},  and it is plausible that
the frequency of hot Jupiters around A-type stars is sufficiently high to explain many ${\lambda}$ Boo stars.

If the wind  from a hot Jupiter provides the material accreted onto a ${\lambda}$ Boo star, then it may be possible
to detect transits or more subtle photometric variations for systems inclined  less than 90$^{\circ}$.  \citet{Balona2013} has analyzed the Kepler light curves for ${\sim}$2000 A-type stars and found  166 candidate hot Jupiters  \citep{Balona2014}.  Further
study is required to determine how many of these systems truly harbor planets.

In our proposal, some of the elements in a wind from hot a Jupiter are  expelled into the interstellar medium by the star's radiation pressure and therefore mimic a stellar wind.  This scenario might explain the observed outflow apparently 
from Sirius \citep{Bertin1995}  which is not otherwise well understood.  Usually the  wind speed  is comparable to the escape velocity, and the modest blueshift of between -20 km s$^{-1}$ to -80 km s$^{-1}$ in the outflow attributed to Sirius may instead be from a planet.

One observational argument against our proposed scenario is that HD 15028 -- the host star of WASP 33, an A-type star with a hot Jupiter-- is classified
as an Am star \citep{Collier-Cameron2010} and not a ${\lambda}$ Boo star.  However, if the wind from the planet is
magnetically funneled onto the host star, then perhaps elemental separations would not occur.  It may be that
only some winds from hot Jupiters lead to the ${\lambda}$ Boo pattern of abundances.   Another observational argument against
our model is that  two ${\lambda}$ Boo stars identified in the Kepler survey \citep{Niemczura2015} are not obvious candidates for harboring
a hot Jupiter \citep{Balona2014}.  However, this sample is tiny and the result inconclusive.   Despite difficulties, the scenario of accretion from hot Jupiters appears to be the least unlikely of all current proposals to explain ${\lambda}$ Boo stars.

Future observational tests of ${\lambda}$ Boo models can be performed with the well defined set of stars provided by \citet{Murphy2015}.

\section{CONCLUSIONS}

We argue that empirical bounds  exclude  interstellar matter as the source of accreted material onto ${\lambda}$ Boo stars.
We then propose two modifications to the popular model to explain the abundance pattern in ${\lambda}$ Boo stars.  First,
we suggest that radiation pressure on individual gas-phase ions might be
 an important mechanism for selecting which elements can be
accreted.  Second, we suggest that the matter acquired by some ${\lambda}$ Boo stars may originate
in the wind from a hot Jupiter.  

This work has been partly supported by the NSF.  I thank B. Hansen and B. Zuckerman for helpful comments.

 \bibliographystyle{apj}

\end{CJK}
\end{document}